# Mission Analysis for the HENON CubeSat Mission to a Large Sun-Earth Distant Retrograde Orbit

Stefano Cicalò[1*], Elisa Maria Alessi[1,6], Lorenzo Provinciali[2],
Paride Amabili[2], Giorgio Saita[2], Davide Calcagno[2],
Maria Federica Marcucci[3], Monica Laurenza[3], Gaetano Zimbardo[4],
Simone Landi[5], Roger Walker[7], Michael Khan[8]

[1*]Space Dynamics Services s.r.l., Cascina (PI), Italy.
[2]Argotec s.r.l., San Mauro Torinese (TO), Italy.
[3]INAF-Istituto di Astrofisica e Planetologia Spaziali, Roma, Italy.
[4]University of Calabria, Rende (CS), Italy.
[5]University of Florence, Sesto Fiorentino (FI), Italy.
[6]IMATI-CNR, Milano, Italy.
[7]ESA-ESTEC, Noordwijk, The Netherlands.
[8]ESA-ESOC, Darmstadt, Germany.

*Corresponding author(s). E-mail(s): cicalo@spacedys.com;

**Abstract**

The HEliospheric pioNeer for sOlar and interplanetary threats defeNce (HENON) mission is a CubeSat Space Weather mission, designed to operate in a Sun-Earth Distant Retrograde Orbit (DRO) at more than 10 million km from the Earth. HENON will embark payloads tailored for Space Weather (SWE) observations, i.e., a high-resolution energetic particle radiation monitor, a faraday cup, and a magnetometer enabling it to provide quasi-real-time monitoring of the interplanetary conditions in deep space. HENON has many important goals, such as demonstrating CubeSat capabilities in deep space, including long-duration electric propulsion with periodic telemetry and command, and robust attitude control for deep-space operations. It will pave the way for a future fleet of spacecraft on DROs, providing continuous near real-time measurements for SWE forecasting. This paper focuses on the mission analysis performed for phase A/B, with the main goal of defining a baseline transfer trajectory to a heliocentric DRO in co-orbital motion with the Earth. The proposed transfer leverages a rideshare opportunity on a mission escaping Earth's gravity field, most likely one headed toward the Sun–Earth $L_2$ region, and relies exclusively on on-board electric propulsion to reach deep space, making it a pioneering demonstration of this approach and the technology. Under appropriate assumptions on the electric propulsion system performances, s/c mass and propellant budget, it will be shown that the HENON target DRO can be reached in about 1 year, taking into account also periodic interruptions of thrusting to allow for Telemetry, Tracking and Command.

# 1 Introduction

The HEliospheric pioNeer for sOlar and interplanetary threats defeNce (HENON) mission is a technology demonstrator for Space Weather (SWE) monitoring, designed to operate a 12U-CubeSat in a Sun-Earth Distant Retrograde Orbit (DRO), at more than 10 million km from the Earth (Marcucci, et al., 2022; Provinciali, et al., 2024). The HENON mission phases A/B have been conducted by a consortium led by Argotec, with INAF, being the scientific PI, University of Calabria, University of Florence and SpaceDyS, under an ESA General Support Technology Programme (GSTP) contract through the financial support of the Italian Space Agency (ASI).

HENON will embark payloads tailored for Space Weather observations, and will exploit the specific type of orbits called Distant Retrograde Orbits, which were studied in detail by M. Hénon in the late sixties in the Hill's approximation (Hénon, 1969).

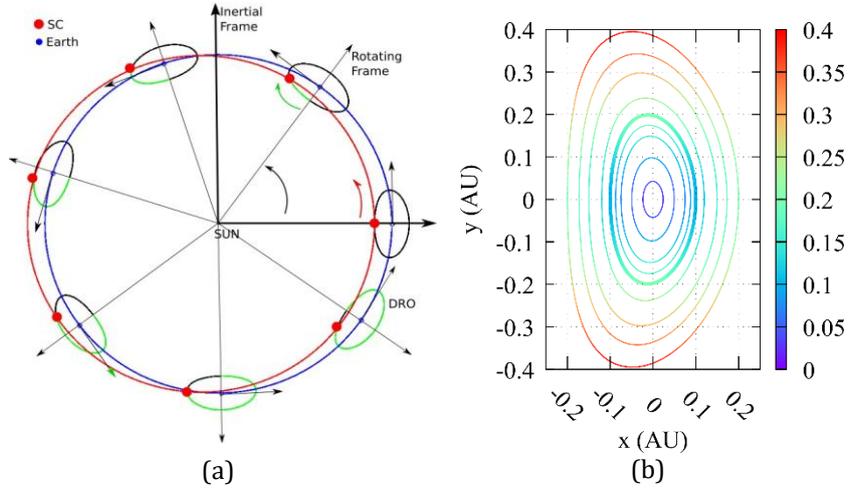

Figure 1 – (a) Concept of Distant Retrograde Orbit in the Sun-Earth system. In the synodic frame, the orbit looks like a retrograde ellipse with respect to the Earth. (b) Distant Retrograde Orbits seen in the rotating Sun-Earth system centered at the Earth. The colorbar represents the distance from the Earth along the orbit, in AU.

DROs correspond to the so-called family $f$ of symmetric periodic orbits, first derived in the framework of the Restricted Three-Body Problem Hill's approximation, when the mass parameter tends to 0 (Jackson, 1913; Hénon, 1969). In the inertial reference system, they orbit the primary (i.e., the Sun) in a 1:1 mean motion resonance with the secondary (i.e., the Earth). Seen in the synodic reference system, these orbits appear to orbit the planet in retrograde motion (see Figure 1). Orbits of the family can be computed for a large range of distances with respect to the planet. In particular there exist orbits well beyond the Hill sphere (i.e., the Lagrangian points $L_1$ and $L_2$) and orbits well inside it (Voyatzis & Antoniadou, 2018). For the HENON mission, this means that the period of revolution around the Sun is of about 1 year and the nominal orbit is chosen such that the spacecraft remains for a long time at a much larger distance than Sun-Earth Lagrangian point $L_1$ (SEL1). This orbital configuration offers highly favorable conditions for real-time SWE monitoring, while also ensuring stable relative orbital motion. A similar mission concept was proposed in the late 1990s with the Diamond mission (Cyr, et al., 2000), which envisioned a constellation of four spacecraft equally spaced along the same DRO at a minimum distance of 0.1 AU from Earth. This arrangement would enable continuous solar wind monitoring, with at least one spacecraft always positioned near the key observational region along the Sun-Earth line. Although the Diamond mission was never implemented, its analysis clearly demonstrated the advantages of using DROs for solar radiation monitoring, benefits that remain highly relevant for HENON. In this context, HENON is designed as a single-spacecraft, low-cost demonstration mission. While it will provide only partial temporal coverage over the course of a year, its purpose is to validate the concept and lay the foundation for a future, more comprehensive constellation-based SWE monitoring system.

## 1.1 Scientific rationale

The wealth of physical processes that occur on the Sun drive various SWE phenomena which can affect the physical state of the interplanetary space and planetary environments (Laurenza, et al., 2023). One of the most dangerous phenomena is Solar Energetic Particles (SEPs), emitted during solar eruptions. SEPs constitute one of the main components of the charged particle radiation in space and near-Earth environment, along with the background of Galactic Cosmic Rays (GCRs). Other harmful events include Interplanetary Perturbations (IP) having their origin on the Sun, such as Interplanetary Coronal Mass Ejections (ICMEs), and solar wind High-Speed Streams (HSSs), that can interact with Earth's magnetosphere-ionosphere system and cause geomagnetic storms and substorms.

The aforementioned SWE phenomena can severely affect both technological systems and human health. For this reason, SWE monitoring, forecasting and mitigating impacts on geospace, spacecraft, astronauts and future crewed missions beyond Earth's orbit have become primary goals of ESA, NASA, ESF, NSF, NOAA, and other federal space agencies.

Given the high potential risks deriving from SWE, it is paramount to extend the forecasting time horizon. In this respect, DROs offer the opportunity for a leap in our capabilities of predicting SWE events. Quasi-real time monitoring of the interplanetary conditions by a spacecraft located at 0.1 AU upstream of the Earth toward the Sun could increase the warning time for geoeffective interplanetary perturbations by an order of magnitude with respect to one located at the SEL1 point. Specifically, a forecast system for SEPs at 0.1 AU would not actually provide a significant improvement in the lead time with respect to SEL1, considering the particle fast speed. As a matter of fact, the primary obstacle to providing reliable warnings of impending SEP events is the rapid and accurate determination of precursor quantities (e.g., the associated flare peak flux, CME speed), given that the lowest-energy protons of interest (~10 MeV) take approximately one hour to propagate from the Sun to the Earth (Laurenza, et al., 2018). In the framework of HENON, a novel SEP forecasting model has been developed by INAF (Stumpo, et al., 2025) which will use on board measurements in KR1 to provide timely and reliable SEP alerts. In particular, the SEP forecasting model exploits the basic idea that quasi-relativistic electrons propagate much faster than energetic protons due to the mass difference. Therefore, because of their different travel time, it is possible to obtain from the quasi-relativistic electrons (that will be also measured by HENON) indirect information about the properties of the energetic proton flux, which is the geoeffective quantity that needs to be predicted. Such an SEP forecasting model will provide a warning about one hour in advance to the SEP arrival at the Earth.

As far as IP magnetic perturbations are concerned, such ICME and HSSs, a quality leap in the forecasting horizon can be achieved by exploiting the HENON orbit. For instance, ICMEs can propagate at velocity more than 1000 km/s and can be revealed few hours (3-5) before their impact on Earth's magnetosphere, compared to about 20 minutes at SEL1. Two types of techniques have been developed by INAF that will use HENON measurements in KR1 for the generation of alerts for geoeffective IP perturbations and for the geomagnetic storms as measured through the Sym-H index. The first technique is based on the overcome of thresholds of specific interplanetary parameters, such as the solar wind velocity and interplanetary magnetic field intensity and components, that past studies demonstrated to be critical for the geoeffectiveness of ICMEs (Gonzalez & Tsurutani, 1987). The second technique uses an Artificial Neural Network. The EDDA (Empirical Dst Data Algorithm) algorithm (Pallocchia, et al., 2006), using only magnetic field data, has been adapted to predict the SYM-H index 1 hour ahead every 20 minutes.

The HENON payload comprises a radiation monitor for the measurement of energetic electrons and protons, two Faraday Cups for the measurement of the solar wind density, velocity and temperature, and a magnetoresistive magnetometer for the measurement of the Interplanetary Magnetic Field (IMF). Such a payload can provide a complete characterization of the properties of the most important SWE phenomena, enabling the early on-board computation and near real time transmission to Earth of alerts for the related events.

The DRO at approximately 0.1 AU from the Earth also allows significant scientific advances in SWE science, especially to: study the radial evolution of ICME (Zhang, et al., 2021) and solar wind turbulence (Klein, et al., 2019), including the large-scale turbulent eddies in the solar wind (in combination with data from SEL1 spacecraft). Moreover, novel data in a new, unique, and unexplored vantage orbit can provide new insights on acceleration and transport of energetic particles (Dresing, et al., 2023) and on the extension of the night side magnetosphere, in analogy with what was studied in the case of Jupiter by Lepping et al. (1983).

The operational orbit is divided in two regions, namely Key Region 1 (KR1) and Key Region 2 (KR2), where different scientific objectives are pursued (Figure *2*). Specifically, HENON will fly in KR1 for ~88 days. For the whole period, HENON and the Earth have their nominal magnetic footpoint within 20° in heliocentric longitude from each other, i.e., they can be considered to be magnetically well-connected (Nolte & Roelof, 1973). This implies that no significant longitudinal gradient in SEP fluxes is expected between HENON and the Earth and therefore a reliable forecast of SEP events at the Earth can be done from observations in the KR1. On the other hand, IP detected in the KR1 could be different from those detected at the Earth when the angle between the Sun-Earth line and the HENON-Earth line is larger than about 45°. Dedicated simulations using the ENLIL model for ICME propagation (https://ccmc.gsfc.nasa.gov/models/ENLIL~2.8f) indicate that these angular differences have a limited effect on the validity of geomagnetic storm alerts issued based on KR1 observations.

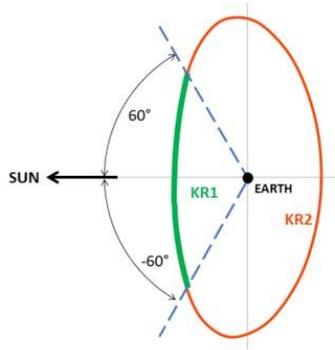

Figure 2 - Key Region 1 (KR1), defined as portion of the operational DRO for the HENON mission where the Sun-Earth-spacecraft angle is less than 60°, and Key Region 2 (KR2).

## 1.2 Trajectory design rationale

Regarding the transfer strategy to reach such large DRO, the general framework in Cyr et al. (2000) assumes a dedicated launch for a spacecraft (s/c hereinafter) with estimated payloads of hundreds of kg, which can provide an additional $\Delta V$ by on-board propulsion for target orbit injection, as suggested in Ocampo & Rosborough (1999). Conversely, HENON is a pioneer demonstration mission, to be realized with a reduced budget, typical of CubeSat missions, and with low-thrust Solar Electric Propulsion as main propulsion system. The HENON 12U Cubesat platform is designed and manufactured by Argotec (Figure *3*). It is expected to have an approximate mass of about 30 kg, equipped with a payload suite consisting of the three instruments aforementioned.

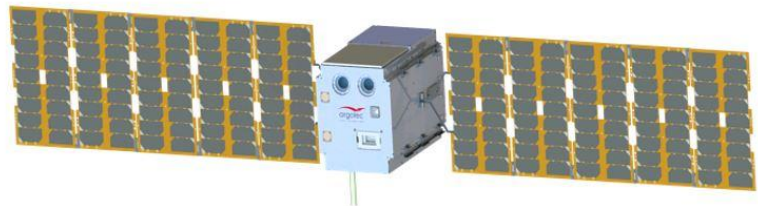

Figure 3 - Rendering of the HENON platform (frontal view).

The payloads are used in KR1 to provide a demonstration of the alert service in near-real time, and in KR2 to collect measurements to better characterize the radiative environment. The satellite avionics possess high radiation tolerance, which in turn is an important limiting factor on the mission duration. The satellite is equipped with a Power Conditioning and Distribution Unit (PCDU) capable of handling as input the almost 200 W of power required to power the platform in the most power-demanding situations, achieved through a 2-wing solar panel configuration with 5 panels per wing, equipped with Solar Array Drive Assembly (SADA) mechanism. The s/c includes also a deep space X-Band transponder capable of reaching 15 W of RF power, to establish a link with Ground even in emergency situations.

The key subsystem, enabler of the interplanetary transfer, is an electric thruster based on Radiofrequency Ion Thruster (RIT) technology, whose performances are described in Section 2.2. It includes, in addition of the electric RIT propulsion system, a Thruster Pointing Mechanism (TPM) and an integrated Reaction Control System (RCS) requested for the on-board wheels off-loading. Completing the satellite bus is a four-wheels ADCS configuration, one of which is used to provide redundancy in the event of failure. The attitude determination capabilities are provided by a high-performance star tracker and a set of sun sensors.

This paper focuses on the Mission Analysis main challenges and results, in particular regarding the optimization of low-thrust transfers to a range of target DROs, in order to find feasible solutions in terms of transfer time and propellant consumption. The strategy adopted is based on the following two steps.

In a first step, a fixed launch injection condition is selected. We use a representative scenario based on the trajectory of a Sun–Earth $L_2$ (SEL2) mission, such as the James Webb Space Telescope (JWST), to model a realistic rideshare opportunity. Although the mission's target is a heliocentric DRO, not SEL2, such trajectories provide suitable escape conditions. While similar strategies could apply to SEL1 missions, SEL2 trajectories were considered due to the higher likelihood of upcoming rideshare opportunities in that direction. Starting from the injection, two main types of maneuvering strategies are identified in order to let the s/c exit the Earth's Sphere Of Influence (SOI) and reach deep space. No optimization procedure is applied during this first step, but two orbital conditions favorably exiting the SOI either from the vicinity of SEL2 or SEL1 are defined.

In a second step, an optimization method is set up to obtain the low-thrust control profile to continue the interplanetary transfer and reach a target DRO. To this aim, a detailed review of possible available methods is provided by Morante et al. (2021). In the present work we adopted a direct collocation method, along with a high-fidelity propagation model, and a Sequential Quadratic Programming (SQP) method named Vf13ad, both available within the NASA *General Mission Analysis Tool* (GMAT, https://documentation.help/GMAT). The corresponding optimal control problem is thus transcribed in a Non-Linear Programming (NLP) constrained problem, as typically done for similar trajectory design analysis (Yam, et al., 2011). The adopted strategy shares many features in common with low-thrust trajectory optimization for planetary or asteroids rendezvous (Mereta & Izzo, 2018), and with low-thrust transfers between DRO and Libration Point Orbits in the Earth-Moon system (Parrish, et al., 2016; Pritchett, et al., 2017). Elegant indirect methods for similar problems are adopted in Topputo et al. (2021), and Zhang et al. (2015). No available literature discusses in details low-thrust transfers to DRO in the Sun-Earth system.

The main assumptions and constraints at the basis of the analysis during HENON phase A/B are presented in Section 2. The transfer design method definition and results are described in Section 3. Conclusions and next steps are given in Section 4.

## 2  Main Assumptions and Constraints

The prime issue underlying the mission analysis process is the selection of the target DRO size. Assuming a circular Earth orbit, DROs with eccentricities between 0.05 and 0.2 reach a minimum distance from the Earth between 0.05 and 0.2 AU. Higher eccentricities are preferable for the scientific purposes, since they would allow for a larger anticipation of SWE alerts, but they generally require a larger $\Delta V$ to be reached from the vicinity of the Earth. Although less popular than the orbits at SEL1 and SEL2, in literature there exist works that analyze how to achieve a transfer to a DRO from Earth, both in the Sun-Earth and in the Earth-Moon system. If we focus on the Sun-Earth system, Ocampo & Rosborough (1999), Demeyer & Gurfil (2007), Scott & Spencer (2010), and Colombo et al. (2014) offer an overview of the most relevant solutions. All of these studies adopt as dynamical model the Planar Circular Restricted Three-Body Problem (PCR3BP) (Szebehely, 1967), and provide reference values of $\Delta V$s for impulsive transfers from Low Earth Orbit (LEO) to large DROs. For DROs characterized by a minimum distance from Earth between 0.05 and 0.2 AU the cost varies from 3.5 and 4.7 km/s, with transfer times of the order of 8-9 months. These $\Delta V$ values refer to impulsive two-burn transfers from LEO, typically assuming that the initial impulse is provided by a dedicated launch vehicle, and include both the eccentricity increase and phasing requirements relative to Earth. They are highly consistent with the $\Delta V$ theoretically needed to achieve such eccentricities via radial escapes. Under the standard two-body energy formulation, assuming an escape along the radial Sun–Earth line and negligible tangential velocity, the heliocentric eccentricity e is given by:

$$e = \sqrt{\frac{a\, V_\infty^2}{\mu}}$$

where a≈1AU, $V_\infty$ is hyperbolic excess velocity, and μ is the Sun's gravitational parameter. This indicates that the majority of the $\Delta V$ in a transfer to a DRO is intrinsically linked to achieving the required orbital eccentricity. This interpretation suggests that leveraging escapes along the colinear lines (SEL1–SEL2 trajectories) would represent a practical strategy for CubeSat rideshares. More recently, in Parsay & Folta (2022) the problem of transfers to DROs through a rideshare to SEL1 has been explored. That paper also investigates the effects of major perturbations on the stability of the orbits. Even though the size of the considered DROs were too small compared to those of interest for the HENON mission (by a factor of 4-5), the basic framework presented in that paper is a valuable reference.

However, two main challenges emerged. Firstly, a dedicated launch is not expected for HENON due to mission cost constraints, and secondly, the transfer $\Delta V$ would have to be provided either by the s/c itself, or by an external carrier vehicle. A straightforward solution for the first issue is to launch HENON as a secondary payload of a suitable larger mission. Due to the dynamical properties of the target orbit and the favorable radial escapes conditions previously mentioned, the launch of a typical Sun-Earth Lagrangian point ($L_1$ or $L_2$) mission would be appropriate (Perozzi, et al., 2017; Parsay & Folta, 2022). In the latter case, dedicated computations showed that the amount of additional $\Delta V$ to be provided by the s/c to reach a ~ 15 million km DRO would be less than 2 km/s, and around 1 km/s for a DRO of ~ 10 million km (see also Section 3.2). Nevertheless, neither the possibility to provide the $\Delta V$ with on-board chemical propulsion, nor the possibility to use a larger external carrier vehicle turned out to be feasible solutions for HENON.

The most promising solution turned out to be the adoption of a miniaturized onboard Electric Propulsion System unit (EPROP), in analogy with the case of the ESA M-ARGO CubeSat mission (Topputo, et al., 2021), which shares several similarities with HENON. The details on the corresponding adopted model for HENON are given in Section 2.2. For instance, assuming a Specific Impulse (Isp) of 3600s, and a thrust level of 1.7mN, a preliminary computation via the Tsiolkowsky equation results in a nominal consumption of about 1.6 kg of propellant to develop a $\Delta V$ of 2 km/s for a s/c of mass 29 kg, in slightly more than 1 year. Given that the HENON mission requires a substantial increase in heliocentric eccentricity after Earth escape, the distribution of low-thrust $\Delta V$ must be optimized to maximize this effect with minimal fuel expenditure. In this context, it is well established that fuel-optimal low-thrust transfers enhancing orbital eccentricity tend to concentrate thrust arcs near periapsis and apoapsis, with intervening coast arcs. This behavior results from classical optimal control theory and variational analyses (Betts, 2001; Prussing & Conway, 2012) where orbital energy and angular momentum are most efficiently modified at the apsides. The same strategy is also evident in early applied mission designs such as Kawaguchi (2002), which, although centered on trajectories combining Earth gravity assists and continuous electric propulsion within the EDVEGA framework, adopts thrusting patterns that implicitly reflect this optimality structure. It represents an early application of these principles in practical mission synthesis. This behavior will be confirmed by the thrust distribution emerging from our own optimal control solutions, discussed in Section 3.3.

Another important issue to be faced by HENON is the actual CubeSat capability to withstand the radiation environment for a long time, during transfer and operational phases. A transfer duration of less than 1.5 years was mandated together with an operational phase of at least 1 year, as a conservative assumption in view of contingency scenarios that would increase the duration of the mission. Moreover, due to limited power budget available on-board, periodic interruptions of thrusting arcs will be necessary to allow for Telemetry, Tracking and Command (TT&C) sessions, also affecting the transfer duration. For this reason, minimum time solutions have been favored, at the price of considering target DROs with a minimum distance from the Earth of less than 0.1 AU.

## 2.1 Launch Scenario

We assume a realistic launch scenario in which the HENON s/c is deployed as a secondary payload on a mission to SEL2. Similar dynamics apply for transfers to SEL1, as supported by prior analyses (Ocampo & Rosborough, 1999; Demeyer & Gurfil, 2007) and discussed further in Section 3.

Typical injection conditions for such missions include a highly eccentric orbit with a perigee near 200 km and an apogee of approximately $1 \times 10^6$ km, with separation occurring within the first hour after launch, consistent with historical missions such as JWST, Herschel, and Gaia. A few trajectory correction maneuvers are generally required post-separation to correct for launcher dispersion.

For this study, we use the post-separation state vector of the JWST transfer (from the JPL Horizons database) to initialize the HENON transfer trajectory (Table *1*). Similar results have been obtained using conditions from other SEL2 missions, supporting the generality and realism of the chosen initial conditions.

Table 1 – Orbital elements to generate s/c initial conditions at separation from launcher, (JWST launch, data from JPL Horizon database https://ssd.jpl.nasa.gov/horizons)

|  | Epoch | Approximate orbit radius (km) | Approximate Osculating Orbital Elements $(a, e, i, \Omega, \omega, \nu)$ in Earth-MJ2000-Ecliptic (km, deg) |
|---|---|---|---|
| Liftoff | 25-Dec-21 ~12:20UTC | | |
| Separation Epoch | ~12:47UTC | ~7700 | |
| First Ephemeris Available | ~13:01UTC | ~11400 | 519960, 0.987, 23.7, 169.5, 110.3, 80.6 |

## 2.2 Propulsion System Model

The EPROP performance model is based on the ESA M-ARGO mission analysis (as provided by ESA for HENON), and updated from Topputo et al. (2021). It consists of the three polynomial functions (Table 2): the available electric power $P$ as a function of Sun-s/c distance $r_S$, and the thrust $T$ and specific impulse $I_{sp}$ as functions of the available power. The function $P(r_S)$ was provided by Argotec based on expected solar panels capabilities, with imposed limits of 130 W (maximum) and 80 W (minimum), to match EPROP expected performances (Figure *4*). The data provides the fixed performance assumptions underlying all transfer optimizations.

Table 2 – Polynomial functions defining the EPROP performances adopted for HENON.

| | Units ($r_S$ in AU) |
|---|---|
| $P(r_S) = 2471.52 - 6753.83\, r_S + 7634.20\, r_S^2 - 4082.24\, r_S^3 + 850.88\, r_S^4$ | W |
| $T(P) = -1.2343 + 0.026498\, P$ | mN |
| $I_{sp}(P) = -5519.5 + 225.44\, P - 1.8554\, P^2 + 0.005084\, P^3$ | s |

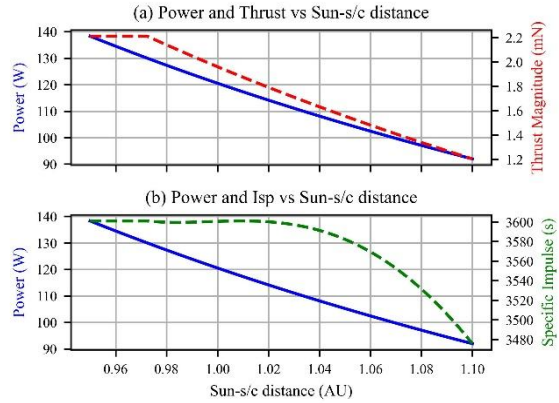

Figure 4 - Performance model regarding the dependency of the electric power available to the EPROP from the Sun-s/c distance (blue curves), and the corresponding values for maximum thrust magnitude and $I_{sp}$.

## 3  HENON Transfer Trajectory Design

The methodology adopted to compute the transfer trajectory for HENON is based on the behavior of the families *a* and *c* of periodic orbits defined in Hénon (1969). These orbits are called Earth-Return-Orbits (ERO) (Demeyer & Gurfil, 2007) and belong to the family of planar Lyapunov Orbits around SEL1/SEL2, as a continuation toward the Earth (see Figure 5-a). Even if they are not stable as the DROs, they have the property of passing very close to the Earth, and of approaching large DROs tangentially. Based on this, the authors in Ocampo & Rosborough (1999) first suggested that optimal transfers to a DRO can be obtained along a slightly adapted half-cycle path of an ERO (see Figure 5-b). They proposed a two-impulse transfer: starting from a low parking LEO, the first $\Delta V$ injects the s/c in a suitable ERO-like trajectory which arrives tangentially to a large target DRO, then a second $\Delta V$ injects the s/c into the target DRO with the correct relative velocity.

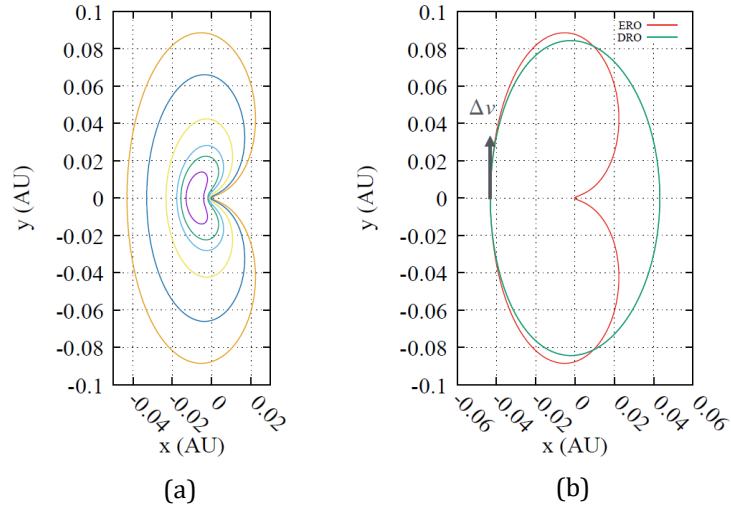

Figure 5 – (a) Family of planar Lyapunov orbits stemming from L1 in the Sun-Earth system, also including the EROs. (b) Sketch of the transfer from Earth to a DRO through an ERO.

This two-impulse transfer strategy is the basis for the design of the HENON transfer. The goal is for the s/c to follow a proper ERO-like trajectory with a continuous low-thrust maneuver instead of impulsive ones, and to finally approach the DRO tangentially at some point. No impulsive maneuvers are included in the final optimized solution, while a simplified two-impulse transfer model can be used as an initial guess for the numerical optimization process. The final trajectory is the result of solving the complete optimal control problem for continuous thrusting. The main constraints and corresponding motivation for the design are summarized in Table 3. The detailed mechanism by which the EPROP $\Delta V$ is distributed along the trajectory, highlighting how thrust is preferentially applied in specific orbital regions to minimize propellant, is analyzed in Section 3.3.

Although techniques such as Earth swing-bys can enhance $\Delta V$ efficiency (Kawaguchi, 2002), these strategies were not considered in this study. The limited transfer time of 1.5 years makes the inclusion of such maneuvers impractical. Moreover, for small s/c such as CubeSats, gravity assist strategies introduce additional complexity and operational risk, particularly in terms of trajectory control, navigation, and timing. Therefore, this study focuses on direct, low-thrust transfers without gravitational assist maneuvers.

Table 3 – High level constraints for the HENON trajectory design.

| To reach a target DRO with a minimum distance from Earth in a range between $10 \times 10^6$ km and $15 \times 10^6$ km | In order to stay at a minimum distance of ~0.1AU from the Earth, and to evaluate a possible trade-off with smaller DROs. |
|---|---|
| Total transfer time < 1.5 y | In order to reduce the exposure of the s/c to radiation environment, considering 1 y of operational phase plus margins. |

| Total propellant consumption < 2 kg | Expected size of the miniaturized EPROP tank. |
| Initial wet mass 29 kg | Current mass budget of the HENON s/c. |

The whole transfer is split in two main phases:

- **Geocentric Phase**: from separation from launcher to the edge of the Earth's Sphere of Influence (SOI);
- **Heliocentric Phase**: from the edge of the Earth's SOI to the target DRO.

The formulation of the two phases and the trajectory optimization strategy are described in Sections 3.1 and 3.2. The numerical computations were performed by means of the NASA *General Mission Analysis Tool* (GMAT), Version R2020a, with mathematical specifications for methods and models reported in NASA (2020). The dynamical model adopted for the numerical propagation during these phases is not limited to the PCR3BP approximation but contains all the relevant perturbations summarized in Table 5. The integration method is the GMAT available *RungeKutta89*, an adaptive step, ninth order Runge-Kutta integrator with eighth order error control.

## 3.1 Transfer Geocentric Phase

This phase begins with the separation of the CubeSat from the launcher according to the JWST initial conditions (Table 1). The equations of motion have the form:

$$\dot{R} = V$$
$$\dot{V} = -\frac{\mu_E}{R^3}R + A_P + \frac{T_{max}}{m}\hat{P} \qquad (1)$$
$$\dot{m} = -\frac{T_{max}}{I_{sp}\,g_0}$$

where $R, V$ are the geocentric position and velocity of the s/c, $m$ is the s/c mass, $\mu_E$ is the gravitational parameter of the Earth, $A_P$ is the sum of dynamical perturbations included in Table 5, $T_{max}$ is the thrust magnitude applied to the s/c along direction $\hat{P} \equiv \hat{V}$. $I_{sp}$ is the EPROP specific impulse, and $g_0 = 9.8$ m/s$^2$ is the standard gravitational acceleration coefficient.

In this work, the trajectory from s/c separation to the exit of Earth's SOI was not subject to optimization. Instead, we adopted a set of conservative, mission-driven constraints, summarized in Table *4*, and performed forward propagation until the SOI boundary. The first constraint reflects an estimate of platform commissioning time, while the others are intended to prevent close approaches to the Earth or Moon. These assumptions are expected to evolve as the mission design matures, and trajectory optimization will be pursued once more accurate launch conditions become available.

Without propulsion, the s/c would return to Earth after approximately one orbital period (~1 month). To escape the Earth's SOI and reach deep space, the s/c must increase its orbital energy via thrusting in the velocity direction. Two main strategies were identified:
1. **Type 1 exit maneuver** – Thrusting begins a few days after separation, allowing the s/c to reach escape energy and exit the SOI near the SEL2 region (Figure 6a).
2. **Type 2 exit maneuver** – Thrusting starts later, guiding the s/c to apogee near SEL2, then back toward Earth with a progressively increasing perigee, eventually escaping via SEL1 (Figure 6b). This trajectory can be seen as a powered heteroclinic connection between SEL2 and SEL1.

Thus, depending on the time interval $\Delta T_c = t_{start} - t_s$ between the separation epoch $t_s$ and the thrust initiation epoch $t_{start}$, the s/c can reach the SOI boundary at epoch $t_0$ with distinct dynamical conditions relative to Earth, leading to an escape trajectory either toward SEL2 or SEL1. Both strategies are robust against variation in injection conditions, and at least one remains feasible even if the delivered apogee altitude deviates from nominal. This approach provides a flexible and operationally realistic framework for early-phase mission design.

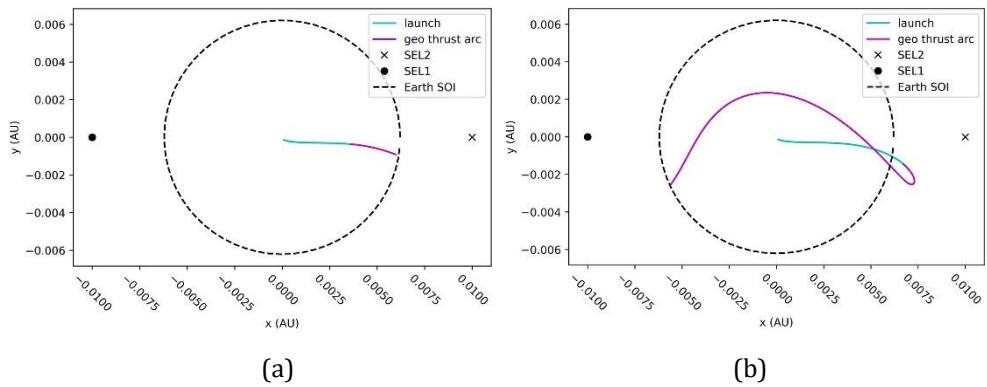

(a)          (b)

Figure 6 - Transfer trajectories during geocentric phases, in the Sun-Earth rotating frame. (a) Exit type 1. (b) Exit type 2.

Table 4 – Parameters under constraints for the geocentric transfer phase and propagated orbital state at the edge of the Earth's SOI.

| Parameter | Constraint |
|---|---|
| Minimum time span between separation and thrusting start $\Delta T_c$ | $\cong 5$ d |
| Minimum distance to Earth and Moon over Type 2 exit maneuver | $\cong 4 \times 10^5$ km |
| **Propagated orbital states** [Earth-MJ2000-Ecliptic, km, s] | |

| | | | |
|---|---|---|---|
| Exit type 1: | $\Delta T_c = 5d$, | $t_0 \sim$ 5-Jan-22, | Distance from Earth |
| $R(t_0) \cong (-0.104, 0.825, 0.336) \times 10^6$; $V(t_0) \cong (-0.143, 0.431, 0.141)$. | | | (SOI) $0.929 \times 10^6$ km |
| Exit type 2: | $\Delta T_c = 19d$, | $t_0 \sim$ 18-Mar-22, | Distance from Earth |
| $R(t_0) \cong (0.747, 0.192, 0.518) \times 10^6$; $V(t_0) \cong (0.247, 0.637, 0.390)$. | | | (SOI) $0.929 \times 10^6$ km |

After exiting the SOI, the s/c needs to keep thrusting towards the target DRO, along a control profile determined by a suitable optimization process described in Section 3.2.

Table 5- Dynamical model adopted for the numerical integration of the trajectory during the Geocentric and Heliocentric phases, which includes all the relevant perturbations listed.

| Dynamical Model's component | Description | | Model |
|---|---|---|---|
| | **Geocentric phase** | **Heliocentric phase** | (NASA, 2020) |
| Primary body | Earth | Sun | |
| Static gravity field | EGM-96 10x10 | Point mass | |
| Third body perturbations | Sun, Moon, and planetary | Earth, Moon, and planetary | JPL DE405 |
| Relativistic correction | General relativity | | |
| Radiation pressure | Solar Radiation Pressure (SRP) ($1m^2$ cross-section, Cr=1.8) | | Isotropic |
| Maneuver | Continuous thrust along the velocity of the s/c relative to the Earth | Thrust along the computed control profile | Section 2.2 for performances |

## 3.2 Transfer Heliocentric Phase

During this phase the equations of motion are numerically integrated in a Sun-centric frame, the dynamic equations now have the form:

$$\dot{\boldsymbol{r}} = \boldsymbol{v}$$
$$\dot{\boldsymbol{v}} = \frac{-\mu_\odot}{r^3}\boldsymbol{r} + \boldsymbol{a_P} + \frac{\eta T_{max}}{m}\hat{\boldsymbol{p}} \qquad (2)$$
$$\dot{m} = -\frac{\eta T_{max}}{Isp\ g_0}$$

where $\boldsymbol{r}, \boldsymbol{v}$ are the heliocentric position and velocity of the s/c, $m$ is the s/c mass, $\mu_\odot$ is the gravitational parameter of the Sun, $\boldsymbol{a_P}$ is the sum of dynamical perturbations included in Table 5, $T_{max}$ is the maximum thrust magnitude applied to the s/c along the direction $\hat{\boldsymbol{p}}$, and $\eta$ is the throttle factor ($0 \leq \eta \leq 1$).

The initial state $(r(t_0), v(t_0), m(t_0))$ comes from the propagation of the geocentric transfer phase (Table 4), and depends on the chosen exit strategy. We start with a distance from the Earth at the order of 1 million km, at the edge of the SOI.

The target condition is to reach a suitable position and velocity relative to the Earth in order to orbit along a well-defined periodic DRO in the Sun-Earth system. The only parameter we used to identify the target DRO is its minimum distance $d$ from Earth at perihelium, being it a fundamental parameter for the SWE alert system. Such condition can be defined in the synodic Sun-Earth rotating frame, with position and velocity coordinates $(x, y, z, \dot{x}, \dot{y}, \dot{z})$ relative to the Earth center, with the $x$-axis oriented along the Sun-Earth direction (away from the Sun), and the $y$-axis oriented along the velocity of the Earth. The DRO planar condition and minimum distance $d$ at perihelium requires that:

$$z(t) = \dot{z}(t) = 0; \quad x(t) = -d; \quad \dot{x}(t) = 0; \quad y(t) = 0; \quad (3)$$

where $t$ is a generic perihelium epoch. The last condition on $\dot{y}$ is computed through differential corrections, searching for $\dot{y}(t) = \dot{y}_d$ such that after one orbital period $P$ of the s/c around the Sun we have:

$$z(t+P) = \dot{z}(t+P) = 0; \quad x(t+P) = -d; \quad \dot{x}(t+P) = 0; \quad y(t+P) = 0. \quad (4)$$

The approach is analogous to what is usually done for computing other families of orbits, such as halo orbits (Breakwell & Brown, 1979). The existence of such solutions in the PCR3BP is guaranteed by the theory, and the high stability of the periodic orbits allows to find solutions also in presence of dynamical perturbations. The only remarkable difference that we found is that, since the eccentricity of the Earth's orbit lets the Sun-Earth distance to vary during the year of about $\pm 2 \times 10^6$ km with respect to 1 AU, the particular value of $\dot{y}(t) = \dot{y}_d$ also depends on the specific epoch of perihelium $t$. We thus computed a table of values $\dot{y}_{d,t}$ over a suitable grid of times $t$ in an interval of years after $t_0$, and distances $d$ in a selected interval between $d_{min} = 0.07$ AU and $d_{max} = 0.1$ AU, in order to allow for interpolation of the required condition at a later stage.

If $t_f$ is the final epoch of transfer, we can check if the target DRO is reached by propagating $(r(t_f), v(t_f))$ (forward or backward) until an epoch $t_p$ such that $(y(t_p) = 0) \& (x(t_p) < 0)$, and verifying the condition:

$$z(t_p) = \dot{z}(t_p) = 0; \; x(t_p) = -d; \; \dot{x}(t_p) = 0; \; \dot{y}(t_p) = \dot{y}_{d,t_p} \quad (5)$$

within an acceptable tolerance. Typically, 1000 km in position and 1 m/s in velocity turned out to be sufficient to obtain a DRO stable over 100 y.

The optimization of this phase of the transfer trajectory is obtained by means of a numerical direct collocation approach, using the nonlinearly constrained parameter optimization routine VF13a (Harwell Subroutine Library, https://www.thinksysinc.com/downloads.html) available for GMAT. The VF13ad optimizer is a Sequential Quadratic Programming based Nonlinear Programming solver. In the analysis presented here, priority has been given to the possibility to easily include perturbations and handle complex constraints, to obtain accurate solutions from a dynamical point of view, at the price of longer computational times. In addition to that, the graphical and script-based user-friendly interface of GMAT allows to quickly set up the optimization problem without the need to implement and test new software.

The methodology follows a direct collocation technique for finding low-thrust optimal solutions in interplanetary rendezvous problems (Yam, et al., 2011; Mereta & Izzo, 2018), and for transfers in the Earth-Moon system (Parrish, et al., 2016; Pritchett, et al., 2017). The main differences in the approach presented here is the use of a fully perturbed dynamical model, and the exploration of transfers to large DROs in the Sun-Earth system.

The optimal control problem can be transcribed in a NLP by splitting the integration time span into equally spaced time segments, imposing a constant thrust vector $\boldsymbol{p} = \eta T_{max} \hat{\boldsymbol{p}}$ acting during each segment. The orientation of the thruster's direction $\hat{\boldsymbol{p}}$ can be defined by standard spherical coordinates $(\alpha, \beta)$ in the local VNB spacecraft frame, whose axes are oriented along the s/c velocity $\hat{\boldsymbol{v}}$, orbit normal $\hat{\boldsymbol{n}}$, and bi-normal $\hat{\boldsymbol{b}}$.

The NLP is then formalized by choosing a suitable number N of subintervals, and splitting the total transfer time $\Delta T = t_f - t_0$ in subintervals of length $\Delta T_N = \Delta T/N$. In our case we chose $N$ to have subintervals of about 10-20 d. For each subinterval $i$ we have a constant throttle factor $0 \leq \eta_i \leq 1$, and a constant thrust direction $(\alpha_i, \beta_i)$. The goal is to propagate the orbital state from $t_0$ to $t_f$ and estimate the thrust profile $(\eta_i, \alpha_i, \beta_i)_{i=1,N}$ satisfying the target constraint (5), and minimizing either the transfer time $\Delta T$ (i.e., to find Time-Optimal TO solutions) or the propellant consumption (i.e., to find Fuel-Optimal FO solutions), see Table 6. The numerical propagation is thus split into $N$ subsequent propagations from $t_i$ to $t_i + \Delta T_N$, for $i = 0,..,N-1$, where the initial conditions of each propagation are the final conditions of the previous one.

Table 6 – Constrained NLP method setup to find TO and FO trajectory solutions.

|  | **Fixed Parameters** | **Variables** | **Non Linear Constraints** | **Objective to Minimize** |
|---|---|---|---|---|
| **TO Solutions** | $(t_0, \boldsymbol{r}(t_0), \boldsymbol{v}(t_0), m(t_0))$ $(\eta_i = 1)_{i=1,N}$ | $(\alpha_i, \beta_i)_{i=1,N}$ $t_f$ | Target constr. (5) | $\Delta T = t_f - t_0$ |
| **FO Solutions** | $(t_0, \boldsymbol{r}(t_0), \boldsymbol{v}(t_0), m(t_0))$ | $(\eta_i, \alpha_i, \beta_i)_{i=1,N}$ $t_f$ | $(0 \leq \eta_i \leq 1)_{i=1,N}$ Target constr. (5) | $-m(t_f)$ |

For TO solutions the thrust is assumed to be always ON. If the throttle factor is further parametrized through, e.g., $\eta_i = (\cos(\xi_i) + 1)/2$, the constraint $0 \leq \eta_i \leq 1$ is always satisfied and can be removed. Coast arcs $(t_i, t_i + \Delta T_N)$ are then characterized by converging close to a constant $\xi_i = \pm\pi$.

The optimization strategy proceeds as follows. A TO trajectory is first computed for the smallest target DRO using a two-impulse transfer as an initial guess. This provides a feasible thrust profile and a reference Time of Flight (TOF). A continuation method is then employed to extend the TO solution over the full range of target DRO sizes, from $d_{min}$ to $d_{max}$, yielding corresponding transfer times $\Delta T_{TO}(d)$. FO trajectories with free final time are subsequently computed for the same DROs initializing from the respective TO solutions. To explore intermediate trade-offs, additional FO solutions with fixed TOF values can be computed by continuation in $\Delta T$, over the interval $[\Delta T_{TO}(d), \Delta T_{FO}(d)]$, producing a parametric family of solutions for each target DRO. This approach provides a coherent framework to quantify how propellant consumption varies with both DRO size and allowable transfer duration.

An initial first guess to compute the TO solutions corresponding to $d = d_{min}$ can be obtained from an adaptation of the two-impulse transfer previously described (Ocampo & Rosborough, 1999). If we consider an Exit Type 1 strategy, starting from $(t_0, \boldsymbol{r}(t_0), \boldsymbol{v}(t_0))$ an impulsive $\Delta \boldsymbol{v}_1$ at $t_0$ can be easily found through differential corrections in order to achieve the condition $x(t_f) = -d$, $\dot{x}(t_f) = 0$, $y(t_f) = 0$, which corresponds to put the s/c on the proper ERO-like orbit and approach the DRO tangentially. Then a second $\Delta \boldsymbol{v}_2$ is computed at $t_f$ in order to reach the correct DRO relative velocity $\dot{y}_{d_{min},t_f}$ in the synodic frame (Figure 5-b). A good first guess for the low-thrust optimization can then be obtained by adapting the setup shown in Table 6, where, for simplicity, we fix the maximum thrust $T_{max} = 2$ mN, and $I_{sp} = 3600$ s, independently of the Sun–s/c distance. This assumption allows for a uniform initialization of the optimization problem, decoupling the first-guess construction from the variation of solar input and enabling consistent convergence across the continuation process. Starting from the following configuration:

$$\boldsymbol{p}_1 = \eta_1 T_{max} \Delta \hat{\boldsymbol{v}}_1, \qquad \eta_1 = \frac{1}{T_{max}} \frac{\Delta m_1}{\Delta T_N} I_{sp} g_0, \qquad \Delta m_1 = m_0 (1 - e^{-|\Delta \boldsymbol{v}_1|/(I_{sp} g_0)}),$$

$$\boldsymbol{p}_N = \eta_N T_{max} \Delta \hat{\boldsymbol{v}}_2, \quad \eta_N = \frac{1}{T_{max}} \frac{\Delta m_2}{\Delta T_N} I_{sp} g_0, \quad \Delta m_2 = (m_0 - \Delta m_1)(1 - e^{-|\Delta \boldsymbol{v}_2|/(I_{sp} g_0)}),$$

$\eta_i = 0$ for $i \neq 1, i \neq N$.

For example, if we have a target DRO with $d \cong 0.07$ AU, the two-impulse results in $\Delta v_1 \cong 1.09$ km/s, $\Delta v_2 \cong 0.2$ km/s, $\Delta T \cong 245\ d$. Although this example is presented in the context of a two-impulse transfer, it serves primarily as a tool for constructing a first-guess trajectory. The s/c begins the transfer from a region near SEL2 with a hyperbolic excess velocity of approximately 0.5 km/s (Table *4*), reducing the total onboard $\Delta V$ required. When interpreted in classical terms, the total energy of the trajectory is consistent with an injection velocity from LEO of ~3.5 km/s, in line with values reported in Demeyer & Gurfil (2007). The NLP constrained problem defined in Table 7 eventually converges to a low-thrust solution with constant thrust of 2 mN, $I_{sp} = 3600$ s, and $\Delta T \cong 280\ d$, which is a good first guess to be used with the full model previously defined in Table *6*. The procedure for Exit Type 2 works similarly.

Table 7 – Constrained NLP method setup to find a first guess for low-thrust trajectory optimization, starting from a two-impulse transfer.

| | Fixed Parameters | Variables | Non Linear Constraints | Objective to Minimize |
|---|---|---|---|---|
| First Guess computation | $(t_0, \boldsymbol{r}(t_0), \boldsymbol{v}(t_0), m(t_0))$ | $(\eta_i, \alpha_i, \beta_i)_{i=1,N}$ $t_f$ | $(\eta_i = 1)_{i=1,N}$ Target constr. (5) | $\Delta T = t_f - t_0$ |

In the continuation process, TO and FO solutions are incrementally computed by varying the target DRO distance $d$ in steps of approximately 0.01 AU.

The overall strategy described here has the advantage to produce high-fidelity solutions, due to the accurate dynamical models and numerical integration method adopted, at the price of long computational time. However, the small dimension of the phase space explored in the optimization problem, and the adoption of parallel computations on multi-core modern computers mitigated this issue. Moreover, among the perturbations considered in Table 5, it turned out that SRP, relativistic corrections, and planetary perturbations other than Earth and Moon have practically negligible effects on the results of the analysis, and can be possibly turned off to save computational time.

## 3.3 Transfer Solutions

We present in this section the most relevant transfer solutions computed, highlighting main advantages and disadvantages in view of possible future developments of the mission.

Figure *7* and Figure *8* illustrate the structure and performance of the low-thrust transfer trajectories for varying target DRO sizes, distinguishing between FO and TO solutions. In both figures, panel (a) shows the FO trajectories with free final time, while panel (b) shows the corresponding TO trajectories for the same DRO targets. The trajectories are plotted in the Sun–Earth rotating frame, allowing clear visualization of the two distinct Earth escape geometries: Figure *7* highlights Exit Type 1 (via the SEL2 region), and Figure *8* highlights Exit Type 2 (via the SEL1 region), with the respective zooms provided in panels (c).

A qualitative inspection of the FO trajectories in panels (a) reveals that the thrust arcs are concentrated near the perihelion and aphelion of the transfer orbits, with long intermediate coasting phases. As expected, this thrusting structure directly reflects the fuel-optimality condition for eccentricity enhancement, as predicted by classical optimal control theory.

Panels (d) show the interpolated evolution of TOF and propellant consumption as functions of the DRO minimum distance from Earth. Both TO and FO data are plotted to reveal the trade-off in consumption and transfer duration, FO solutions yielding lower propellant consumption but longer flight times. They all require less than 1.5 y of transfer time, and less than 2 kg of propellant consumption. In this respect, they provide the basis for selecting baseline mission profiles based on a trade-off between efficiency and duration.

An example of result of the thrust profile for a FO solution for a middle size DRO with $d \cong 0.082\ AU$ is shown in Figure 9. The throttle factor approaches a bang-off-bang profile, as predicted by theory.

As a matter of fact, periodical sessions of communication and ranging are necessary during the entire mission, for tracking and orbit determination. Given that, with the current system design, it is not possible to keep the transponder to work simultaneously with the propulsion system, because it would require too much power. Thus the need to periodically stop the thrusters must be considered, which will increase the transfer time. A comprehensive analysis for an optimal definition of the thruster periodic ON/OFF duration would have to take into account the navigation and orbit accuracy requirements. At this stage, we make general assumptions based on the strategy proposed for M-ARGO and other missions[1]. Such strategy consists of periods of 6 days of continuous thrusting and 1 day of no thrust, allowing for typically 8 hours of communication and ranging. TO transfer trajectories can be recomputed by imposing a constraint on the thrust arcs requiring a 1-day coast period every 6 days, slightly adapting also the geocentric phase to satisfy Table 4 constraints. The corresponding results are shown in Figure 10. The increase of transfer time is noticeable, and Exit Type 2 solutions appear to be favored in terms of both transfer time and consumption, likely because the s/c spends more transfer time closer to the Sun, where the EPROP performs better.

---

[1] ESA private communication

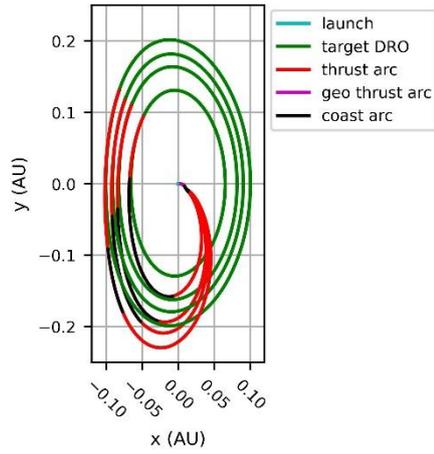
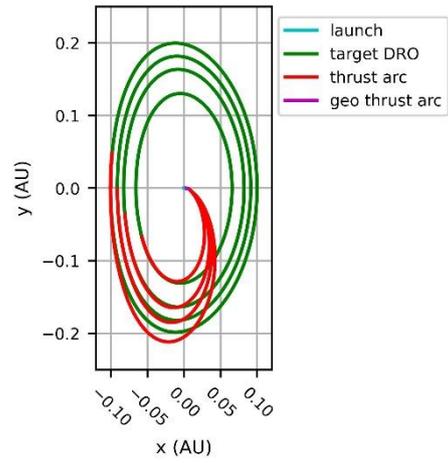

(a)  (b)

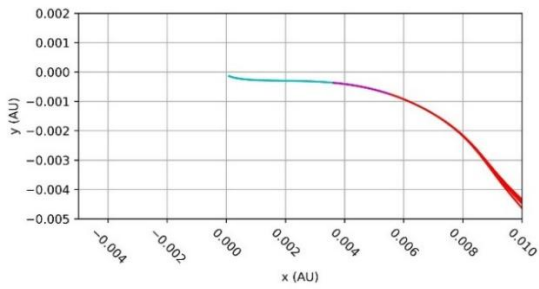
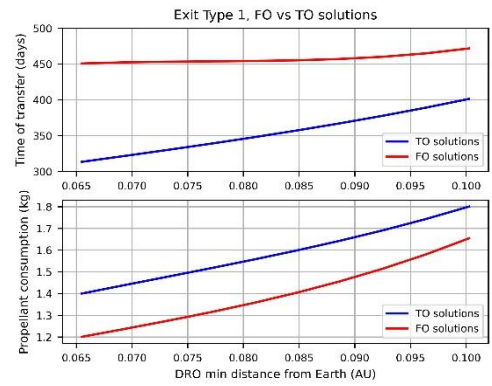

(c)  (d)

Figure 7 – (a) FO low-thrust transfers to DROs of varying size, characterized by their minimum distance from Earth. (b) Corresponding TO low-thrust transfers to the same DROs. (c) Zoomed view of Exit Type 1 trajectories, where the s/c escapes the Earth's SOI via the SEL2 region. (d) Variation of TOF and propellant consumption with DRO size, shown for both FO and TO solutions.

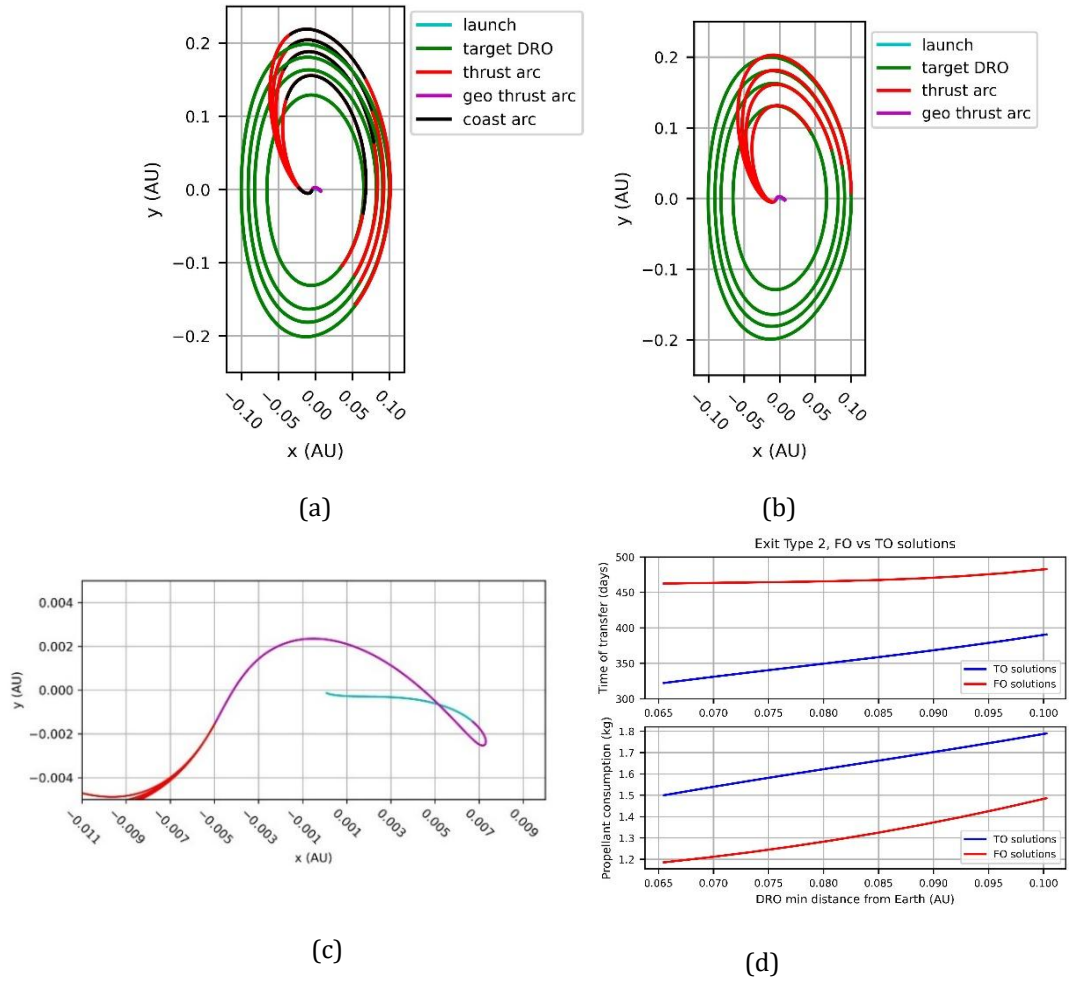

Figure 8 – (a) FO low-thrust transfers to DROs of varying size, characterized by their minimum distance from Earth. (b) Corresponding TO low-thrust transfers to the same DROs. (c) Zoomed view of Exit Type 2 trajectories, where the s/c escapes the Earth's SOI via the SEL1 region. (d) Variation of TOF and propellant consumption with DRO size, shown for both FO and TO solutions.

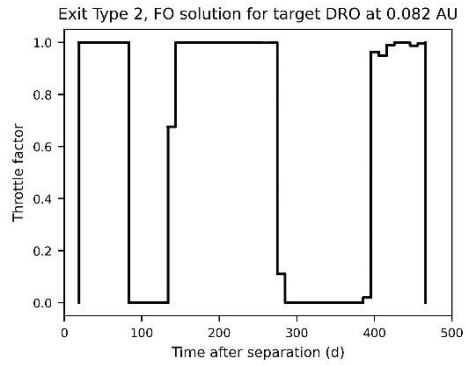

Figure 9 – Example of FO transfer profile to a DRO with minimum Earth distance d=0.082 AU, showing a near bang-off-bang structure. Thrusting is concentrated near perihelion and aphelion, with a coast arc in between, in line with classical low-thrust strategies for modifying orbital eccentricity.

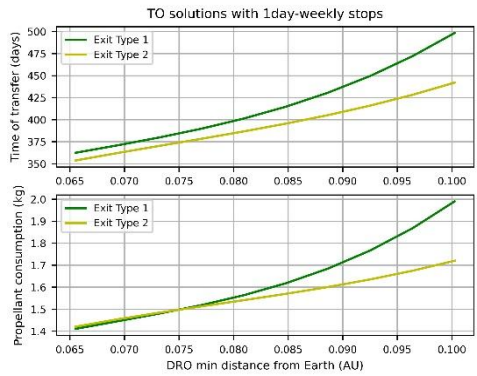

Figure 10 - Interpolated variation of TO transfer duration and propellant consumption with respect to the target DRO size, under the constraint of forced weekly 1-day thruster shutdowns. These coast periods simulate periodic TT&C sessions, during which thrusting is paused due to power limitations. Exit Type 1 solutions (escape via SEL2) and Exit Type 2 solutions (escape via SEL1) are shown. The introduction of periodic coast arcs leads to longer transfer durations, and Exit Type 2 solutions appear more efficient under these constraints.

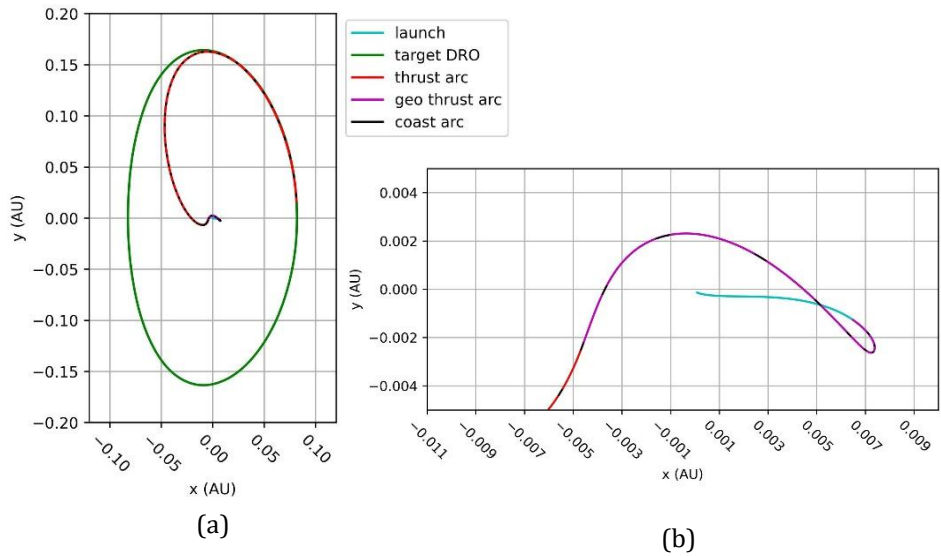

Figure 11 – TO transfer trajectory to a DRO with minimum Earth distance d=0.082 AU, computed under the constraint of weekly 1-day thruster shutdowns for TT&C sessions. (a) Full interplanetary trajectory in the Sun–Earth rotating frame, illustrating an Exit Type 2 departure from the Earth's SOI. (b) Zoomed view of the geocentric transfer phase, showing the early thrust arcs prior to SOI escape.

Moreover, the Exit Type 1 solutions approach the DRO from the side close to the Sun (KR1 region), which is the one where the main SWE experiment has to be performed. Entering the DRO in the middle of the region would not be preferable, since almost one year is necessary for the s/c to enter the KR1 again and be able to perform a full scientific experiment. In this respect, an arrival at the target DRO on the opposite part of the Sun is envisaged, in order to have enough time to prepare and calibrate the scientific instruments before entering the KR1 region. The Exit Type 2 solutions have this behavior.

A final trade-off between transfer time, propellant consumption, and scientific needs has been found in a TO solution with weekly 1d forced thruster stops, with a target DRO size of 0.082 AU, shown in Figure 11.

If the spacecraft is properly inserted into the target orbit, the DRO is a very stable environment, so station-keeping maneuvers are expected to be very small. Moreover, numerical simulations under a full dynamical model have shown that there is not a risk of returning to the Earth within 100 years, which will simplify also the definition of the disposal strategy.

# 4 Conclusions and Next Steps

The mission analysis main challenges and results obtained during the HENON mission phases A/B have been presented. The trajectory design has taken into account realistic mission constraints related to launch scenario, propulsion system, and scientific objectives, as well as the existence of families of periodic solutions that facilitate the goal. High-fidelity dynamical solutions have been generated by means of a direct collocation method, implemented through the GMAT software tool. The final trade-off solution chosen takes about 389 d of transfer time, and ~ 1.55 kg of propellant consumption, to reach a target DRO with a minimum distance from Earth of ~ $0.082\ AU \cong 12.3$ million km, starting from a typical launch scenario of a SEL2 mission. This solution already takes into account periodic weekly interruptions of thrust to allow for telemetry, tracking and command. The trajectory sequence and mission phases presented in this study were developed fully within the operational and programmatic constraints defined for the HENON mission, ensuring the feasibility of the proposed solution for real implementation.

Improvements to the presented analysis will include a proper optimization of the geocentric transfer phase, especially in case of Exit Type 2 strategy. As part of future work, the overall analysis will have to be adapted when more accurate information about the actual launch for the HENON mission will be available, or in case the EPROP performances will be updated.

The next phases of the mission development (i.e., C/D) also foresee the introduction of sources of errors in the analysis, and the definition of the strategies to cope with them. Indeed, the computations and analysis described here assume the capability to navigate the s/c without errors in the propulsion and control systems, and with an arbitrarily good knowledge of the orbital conditions at launch. A complete and reliable error model for the s/c GNC performances will be taken into account. This shall include at least errors due to launcher dispersions, a suitable error model for the propulsion system pointing mechanism and thrust magnitude, and errors in the attitude and orbit determination.

Finally, the possibility of minor or major failures occurring during the mission has also to be considered in a comprehensive analysis. Such contingency scenarios include propulsion outage (due to failures or safe modes), and lower thrust availability (due to power efficiency or thruster underperformance). Their impact on the transfer phase and on the possibility to reach the target DRO will have to be assessed.

**Funding.** The HENON activities have been funded by the Italian Space Agency as part of the ALCOR programme.

**Acknowledgements.** This study has been performed in the framework of the HEliospheric pioNeer for sOlar and interplanetary threats defeNce (HENON) mission Phase A/B. HENON is part of the Italian Space Agency (ASI) program ALCOR and is being developed under the European Space Agency General Support Technology Programme (ESA-GSTP) through the support of the national delegations of Italy (ASI), UK, Finland, and Czech Republic. The authors want to thank G.B. Valsecchi for having strongly supported the start-up of the project. The authors are grateful to the anonymous reviewers, whose comments have helped to improve considerably the paper.

**Ethics declaration.** Not applicable.